\documentclass[prl,preprint,showpacs,preprintnumbers,amsmath,amssymb]{revtex4}
\usepackage{graphicx}% Include figure files
\usepackage{dcolumn}% Align table columns on decimal point
\usepackage{bm}% bold math
\usepackage[mathscr]{eucal}

\begin{document}

% Full title of the paper (Capitalized)
\title{On the unattainability of absolute zero temperature and the Nernst heat theorem}

% Authors, for the paper (add full first names)
\author{Koun Shirai}
\affiliation{%
Nanoscience and Nanotechnology Center, ISIR, Osaka University, 8-1 Mihogaoka, Ibaraki, Osaka 567-0047, Japan
}%

% Abstract (Do not use inserted blank lines, i.e. \\) 
\begin{abstract}
It is sometimes argued that the unattainability of zero temperature is a consequence of the second law of thermodynamics. Historically, the independence of the unattainability of zero temperature from the second law was proven more than 80 years ago, yet this assertion was repeated in the literature. This assertion naturally leads to a doubt that the unattainability of zero temperature is not equivalent to the Nernst heat theorem. The apparent contradiction between the Nernst heat theorem and residual entropy further complicates the problems of the third law. Totally, the validity of the third law seems to lose, giving an impression of somewhat ambiguous hypothesis to it.
The author has recently settled the apparent contradiction between residual entropy and the Nernst heat theorem by refining the statement of the third law.
Based on this refinement, two controversial problems, the independence of the unattainability of zero temperature from the second law and the equivalence of the unattainability with the Nernst heat theorem, have been solved.
% These problems all together bring an impression that the third law is merely an empirical rule which holds under restricted conditions.
% Often claimed view of violation of Nernst heat theorem owing to residual entropy gives support to this suspicion. 
\end{abstract}

\maketitle
%%%%%%%%%%%%%%%%%%%%%%%%%%%%%%%%%%%%%%%%%%
%% Only for the journal Gels: Please place the Experimental Section after the Conclusions

%%%%%%%%%%%%%%%%%%%%%%%%%%%%%%%%%%%%%%%%%%

\section{Introduction}
Although it was established more than 80 years ago, it is strange to find that, even now, the third law of thermodynamics is subject to controversy. 
The problems of the third law in the early days are well described, for example, by Beattie and Oppenheim \cite{Beattie}.
One of the reason for the controversy lies on the manner of the expression.
The law is often stated as

% \begin{Theorem}[Third law: Expression (I)]
\noindent
{\bf Third law: Expression (I)}
{\em The entropy of any system vanishes as temperature approaches zero.}
% \end{Theorem}

\noindent
This is called the Nernst (heat) theorem, although the original expression given by Nernst was slightly different (\cite{Callen}, p.~277). 
% Expression (I) is brief, but because of this briefness, the law was subject to criticisms. 
However, there are many exceptions for Expression (I) due to residual entropies. Glass materials and random alloys are known to have residual entropies. With the progress of material sciences, further exceptions are being found. Exceptions are no longer exception, which render the validity of the third law being restrictive. 

There are other types of problem for the third law. The third law can be restated as 

\noindent
{\bf Third law: Expression (II)}
{\em It is impossible to reach absolute zero temperature.} 

\noindent
This expression is called the unattainability of absolute zero temperature \cite{comment3}. In this case, there is no exception. The contrasting characters of the two expressions naturally arouse a doubt that the two expressions are not equivalent \cite{Hatsopoulos, Hasse,Levine, Landsberg57,Landsberg78,Landsberg97,McNabb17,Masanes17}.
Hasse gave a thorough analysis of the relations between the Nernst theorem and the unattainability of zero temperature, but after all he gave the conclusions in a case-by-case manner \cite{Hasse}. 
Some authors argued that Expression (II) is a consequence of the second law of thermodynamics.
\begin{quote}
{\bf (Statement 1)} The unattainability of zero temperature is deduced from the second law.
\end{quote}
Simon criticized Statement 1 by analyzing the arguments of other authors \cite{Simon51}.
Notwithstanding, objections such as Statement 1 appeared repeatedly in disguised forms. 
The problem of residual entropy further complicates the above questions.

Recently, the author has settled the problem of residual entropy by refining the expression of the third law. By defining the internal constraints accurately, he found a quantitative expression for the third law while acknowledging the existence of residual entropy. This paper is referred to as Paper (I).
Now that the problem of residual entropy has been resolved, we are able to address other problems. In this paper, two controversial issues are solved from the modern viewpoint made in Paper (I).
First one is the independence of Expression (II) from the second law. 
The second one is the equivalence of Expression (II) with (I). 
Although a basic proof was given a long time ago by Simon \cite{Simon27} (see also Fowler and Guggenheim \cite{Fowler-Guggenheim}), 
it is needed to improve the previous methods of proof while considering the problem of residual entropy. The readers are highly encouraged to read Paper (I).

%%%%%%%%%%%%%%%%%%%%%%%%%%%%%%%%%%%%%%%%%%%%%%%%%%
%%%%%%%%%%%%%%%%%%%%%%%%%%%%%%%%%%%%%%%%%%%%%%%%%%
\section{Independence of the second law}
% \section{Cycle versus partial path}
%%%%%%%%%%%%%%%%%%%%%%%%%%%%%%%%%%%%%%%%%%%%%%%%%%
% I like to ask reader's generosity for repeating the proof.
\paragraph{Second law} 

The first issue to be discussed is the independence of the unattainability of zero temperature from the second law.
The main logic underlying Statement 1 is based on the efficiency of the Carnot cycle $\eta_{C} = 1 -T_{l}/T_{h}$, where $T_{h}$ and $T_{l}$ are the temperatures of the hot and cold heat reservers, respectively. When $T_{l}=0$, the efficiency $\eta_{C}$ becomes unity. This is equivalent to claim that the heat received from the hot reserver can be completely transferred to work. 
With this result, it is claimed that reaching zero temperature contradicts the second law.
The second law is sometimes stated in a brief form as
\begin{quote}
{\bf (Statement 2)} Perfect conversion of heat to work is impossible.
\end{quote}
The problem may be rooted in this brief expression of the second law, which prevails over many people's mind without care.
As a matter of fact, we can completely convert heat to work. For example, an isothermal expansion of an ideal gas in a cylinder does. Statement 2 must be replaced with a correct one,

% \begin{Theorem}[Third law: Expression (I)]
\noindent
{\bf The Second Law:} There is no heat {\em engine} that can perform a complete conversion of heat to work.
% \end{Theorem}

\noindent
The author apologizes to write such an easy matter in an original paper, but it is worth emphasizing here, because many erroneous assertions after all originate from this. 
% the author's observation that even leaned people deduce a wrong conclusion owing to this dogma of easily misleading Statement 2.

Let us analyze the problem behind Statement 2. Suppose that an ideal gas fills a cylinder with the initial volume $V_{1}$, contacting a low-temperature heat reserver at $T_{l}$. Let the entropy of the initial state of the gas be $S_{1}$. % This cylinder may constitute a Carnot engine. 
Assume that all the processes are performed in a reversible manner. On contacting a high-temperature heat reserver at $T_{h}$, the heat received by the gas can be completely transmitted to an external device in a work form.
As a result, the volume of the gas is increased to $V_{2}$ and the entropy is increased to $S_{2}$. 
Unfortunately, this process alone is not usable for continuous operations, for the obvious reason that the ever-expanding cylinder cannot be equipped in cars.
Heat {\em cycle} is needed. The working gas must be returned to the initial state of unexpanded volume $V_{1}$ at the low temperature $T_{1}$. Therefore, {\em the necessity of the recovery of the initial state precedes the necessity of heat rejection.} The latter is merely a means for achieving the former.
When $T_{l}=0$, % if the volume of a gas is a continuous function of entropy, $V=V(S)$, 
we can restore the original state $V_{1}$ in an isothermal process without rejecting heat, because $Q=T_{l} \Delta S$ vanishes for any change in entropy $\Delta S$. There is no conflict with the second law.
This was figured out a long time ago \cite{Epstein,Pippard}. Nonetheless, Statement 2 is so appealing that we may be vulnerable to this statement without care.

\paragraph{Significance of heat engines} 
We have to ask why we stick to the Carnot engine for obtaining zero temperature.
The impossibility of the operation of the Carnot engine with $T_{l}=0$ is correct, but this merely states that the engine is incapable of operating with the low-temperature reserver at $T=0$. 
% The zero-temperature world is the one that will be immediately evaporated once it contacts with any finite-temperature world. We have to isolate from any finite-temperature world to maintain its temperature. 
% If we isolate the zero-temperature world from nonzero-temperature worlds, it can survive. 
Maintaining is different from reaching there.

A similar argument holds for the Carnot refrigerator to obtain the zero temperature \cite{McNabb17}. It is correct to state that zero temperature cannot be reached by any refrigerator, how idealized it is. Again, a heat cycle is not necessary for reaching zero temperature, and the use of heat cycle makes the problem even worse. A Carnot refrigerator, by construction, assumes that it is able to cool down adiabatically the working substance to the temperature of the cold reserver. If zero temperature is reachable in this step, no further step of the cycle is required. A single path is the best choice: cycles need to pay a price. 
% The same logic as the discussion of Statement 2 is applied to this problem. 

% Another subtle problem was a question raised by Einstein. According to Simon \cite{Simon51}, Einstein's claim is that, unavoidable heat generation, even though infinitesimally small, by operation of a piston prevents any process from reaching zero temperature. Simon pointed out the difficulties of Einstein's argument, based on the asymptotic behaviors of thermal properties of materials. Their dispute seems pointless, for it is based on the limiting properties of materials as $T \rightarrow 0$. Physics laws must hold independent of material properties. 

% Otherwise, additional (fourth) law would be required. 
% In the following, no specific properties of materials are used.

% \pagebreak
%%%%%%%%%%%%%%%%%%%%%%%%%%%%%%%%%%%%%%%%%%%%%%%%
\section{Path to zero temperature and Irreversibility}
The second issue is the equivalence of Expression (I) to Expression (II).
The author believes that the debates on the equivalence of Expression (I) and (II) ended in the 1940s with a conclusion confirming the equivalence \cite{Simon51}. 
Despite this, the issue is still controversy in the physics community. 
The reason why this is so confusing lies mainly on the treatment of residual entropy. Let us discuss the equivalence in the order from easy to difficult.

\paragraph{Reversible path}
We have seen in the foregoing argument that the use of a single adiabatic path is the best choice for obtaining the zero temperature \cite{adiabatic-change}. A proof of the unattainability of zero temperature will be, therefore, completed by showing that there is no adiabatic path to reach $T=0$. 
Indeed, this is what Fowler and Guggenheim did in their textbook \cite{Fowler-Guggenheim}. 
Here, the author first follows their method for developing theory further.
If an elegant idea of adiabatic accessibility \cite{Lieb99} is used, the proof is completed by a few lines of text \cite{Wreszinski09}.
% $A \not\prec B$
Nevertheless, the method of Fowler and Guggenheim has a merit in the present context, which will be clearer later.

% There is no better way other than adiabatic processes for obtaining a low temperature as ever achieved. Utilizing heat conduction requires a heat reserver, whose temperature is already lower than that of the substance to be cooled.
The best adiabatic processes are reversible ones in terms of the best efficiency.
% ------------------------------------------------------------------------
\begin{figure}[ht!]
\centering
\includegraphics[width=.98 \textwidth]{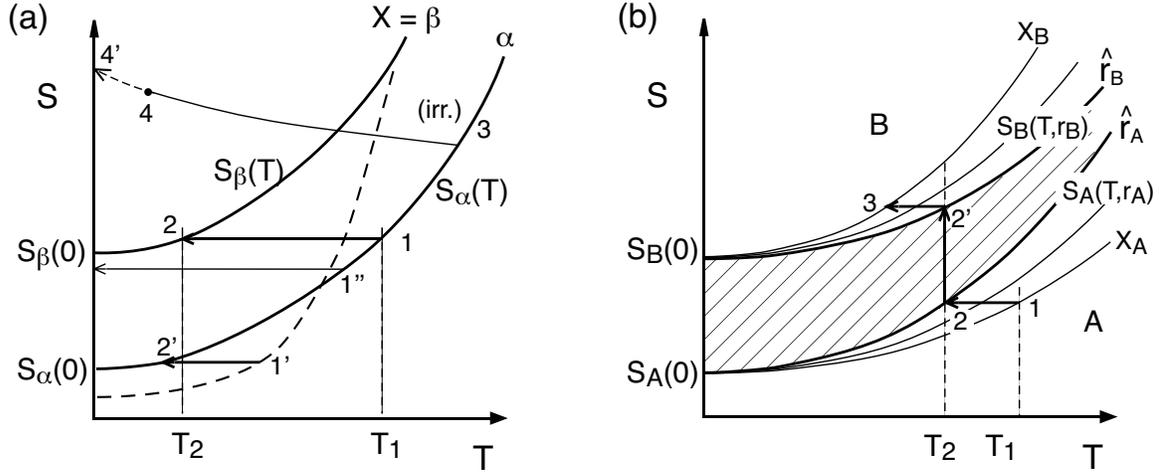}
\caption{
Temperature-entropy diagram for obtaining zero temperature.
(a) Entropies along paths of a constant $X$ all must converge to a single point.
(b) For the case of residual entropy between two systems $A$ and $B$, there is a region inhibited from both systems (hatched region).
}
\label{fig:TS-diagram} 
\end{figure}
Figure \ref{fig:TS-diagram}(a) shows a $T-S$ diagram for a given system $A$ near $T=0$. 
The $S(T)$ curve shows a monotonously increasing function of $T$, which is required by the stability of materials. 
System $A$ has a state variable $X$ other than $T$ and $S$. For example, $X$ can be volume $V$. Along a path of a constant $X=\alpha$, the entropy at an arbitrary $T$ is given by
\begin{equation}
S_{\alpha}(T) = S_{\alpha}(0) + \int_{0}^{T} \frac{C_{X}(T')}{T'} dT'. 
\label{eq:Salpha}
\end{equation}
% where $S_{\alpha}(0)$ is the entropy at $T=0$.
Similarly, the entropy $S_{\beta}(T)$ is given for another value $\beta$ of $X$.
Consider an adiabatic transition from state 1 on $X=\alpha$ to state 2 on $X=\beta$. The respective temperatures are $T_{1}$ and $T_{2}$. Since we are treating reversible processes, the entropy is preserved. 
\begin{equation}
S_{\alpha}(0) + \int_{0}^{T_{1}} \frac{C_{\alpha}(T')}{T'} dT'
= S_{\beta}(0) + \int_{0}^{T_{2}} \frac{C_{\beta}(T')}{T'} dT'.
\label{eq:Salphabeta}
\end{equation}
Suppose that we can reach $T_{2}=0$. Then, we must have 
\begin{equation}
S_{\beta}(0) - S_{\alpha}(0) = \int_{0}^{T_{1}} \frac{C_{\alpha}(T')}{T'} dT'.
\label{eq:Salphabeta1}
\end{equation}
If $S_{\beta}(0) > S_{\alpha}(0)$, we can always find the nonzero solution for $T_{1}$ in Eq.~(\ref{eq:Salphabeta1}), because specific heat is always positive. This result contradicts the unattainability of zero temperature. Hence, it must be $S_{\beta}(0) \leqq S_{\alpha}(0)$.
Now, we can start from a state $1'$ on $X=\beta$ such that $S_{\beta}(1') > S_{\alpha}(0)$, and bring the state towards $2'$ on $X=\beta$ by a reversible adiabatic path. % since the stability condition dictates $C_{X} = T(\partial S/\partial T)_{X}$ being positive. 
A similar argument leads to $S_{\beta}(0) \geqq S_{\alpha}(0)$. These two results lead to only a solution $S_{\beta}(0) = S_{\alpha}(0)$ which can be allowed.

Conversely, if $S_{\beta}(0) = S_{\alpha}(0)$ holds, there should be no reversible adiabatic path $\alpha \rightarrow \beta$ from a finite $T_{1}$ to $T_{2}=0$. For if a path $\alpha(T_{1}) \rightarrow \beta(T_{2}) $ exists, then Eq.~(\ref{eq:Salphabeta}) would lead to a finite $T_{1}$ satisfying
\begin{equation}
S_{\beta}(0) - S_{\alpha}(0) = \int_{0}^{T_{1}} \frac{C_{\alpha}(T')}{T'} dT'.
\label{eq:Salphabeta2}
\end{equation}
This contradicts the initial assumption $S_{\beta}(0) = S_{\alpha}(0)$, because the right-hand side of Eq.~(\ref{eq:Salphabeta2}) is positive. Thus, we have proven the equivalence between Expressions (I) and (II).
Similar proofs, but essentially equivalent, are seen in the literatures \cite{Pippard, Guggenheim, Wilson, Wilks}.

Readers may be concerned, in Fig.~\ref{fig:TS-diagram}(a), that there is a region wherein no isentropic path connecting $\alpha$ and $\beta$ exists, {\it e.g}, an isentropic path starting from state $1''$ on $X=\alpha$ to any state on $X=\beta$. However, it is physically sound to assume that there is continuity in the states of $A$ between $\alpha$ and $\beta$. We can always find a constant line $X=\beta'$, which has the intersection $2''$ between the isentropic path starting from $1''$ and a line $X=\beta'$. % No problem occurs.

% \paragraph{Objections}
\paragraph{Irreversible path}
There is an objection to the above argument. Hasse claims that the above proof using both ways of discrimination, $S_{\beta}(0) \leqq S_{\alpha}(0)$ and $S_{\beta}(0) \geqq S_{\alpha}(0)$, does not hold when an irreversible path is considered \cite{Hasse}. 
Let us consider an irreversible adiabatic path starting, for example, from $3$ on $X=\alpha$ of Fig.~\ref{fig:TS-diagram}(a). Suppose that, at $T=0$, there is such a state $4'$ on a curve $X=\beta'$ whose entropy is larger than $S_{\alpha}(0)$.
% on a curve  has a larger entropy, $S_{\beta'}(0) > S_{\alpha}(0)$. 
The transition $3 \rightarrow 4'$ does not conflict with the second law. 
The problem of his argument is that it misses the fact that the adiabatic process ceases at some point $T_{4}$ between $T_{3}$ and $T=0$, because the temperature that can be reached by irreversible processes starting from a given state is always higher than the temperature that would be reached by the reversible process starting from the same state. 
This shows directly the unattainability of zero temperature: the above proof by the both ways of discrimination is unnecessary. 
% This shows that the judgement of accessibility by the entropy difference between the terminal states only is risky.
% This is not exception in the argument using adiabatic accessibility \cite{Wreszinski09}.
% use of the relationship, $A(\alpha,T_{1}) \prec B(\beta, 0)$ if $S_{A}(\alpha,T_{1}) \leq S_{B}(\beta, 0)$, requires care \cite{Wreszinski09}. Here, we see only the terminal points, $A(\alpha,T_{1})$ and $B(\beta, 0)$. We are liable to fall in a pitfall behind this brief expression. When there is a residual entropy $S_{0}^{AB}$ between $A$ and $B$, we should see the process between the terminal points. The adiabatic accessibility is rewritten as $A(\alpha,T_{1}) \prec B(\beta, 0)$ if $S_{A}(\alpha,T_{1}) > S_{B}(\beta, 0)+ S_{0}^{AB}$.
The arguments hitherto discussed were already given by Fowler and Guggenheim \cite{Fowler-Guggenheim} and others.

% Formal argument using adiabatic accessibility is liable to miss \cite{Wreskinsk09}.
\paragraph{Case of residual entropy}
Now, we are in a position to discuss the special case of irreversible processes in which residual entropy is involved. In this case, there indeed exists a state having a finite entropy at $T=0$, as shown in Fig.~\ref{fig:TS-diagram}(b). In a review paper by Landsberg, by using a similar figure, while claiming that the unattainability of zero temperature holds within each of branches $A$ and $B$, he reserved judgement about the transition between $A$ and $B$ \cite{Landsberg78}. Hatsopoulos and Keenan claimed that there is no proof that excludes the possibility of an irreversible adiabatic process of this type (\cite{Hatsopoulos}, p.~29 in Forward). 

As discussed in paper (I), a residual entropy arises between two systems that are separated by a special internal constraint expressed by a frozen coordinate. In the thermodynamic context given by Gyftopoulos and Beretta \cite{Gyftopoulos}, even for the same material, if two states have different internal constraints, we have to regard two ``states" as different systems.
For example, ice crystals having ordered and disordered structures belong to different systems, despite consisting of the same water molecules.

Figure \ref{fig:TS-diagram}(b) shows a $T-S$ diagram of systems $A$ and $B$ when a residual entropy is present between them. Two systems $A$ and $B$ are separated by a frozen coordinate $r$: they are called belonging to different thermodynamic classes in Paper (I).
The value is fixed at $\hat{r}_{A}$ for $A$ and at $\hat{r}_{B}$ for $B$, leaving a finite difference $\Delta r^{AB} = \hat{r}_{B} - \hat{r}_{A}$. 
Since, by an appropriate transformation, the total entropy can be factorized as $S = \sum_{j} s_{j}(q_{j})$ (Eq.~7 in Paper I), it is permissible to use a simple functional form $S(T,\hat{r})$ in the following argument.
% $S_{A}(T,X_{A},\hat{r}_{A})$, and likewise $S_{B}(T,X_{B},\hat{r}_{B})$. 
% In the figure, thermodynamic coordinates $X_{A}$ and $X_{B}$ other than $T$ and $r$ appear for $A$ and $B$, respectively, which may or may not be the same coordinates. 
% In the following, $X$'s are omitted from the arguments, but readers must bear the scheme of Fig.~\ref{fig:TS-diagram}(b) in mind.
Let us assume $S_{B}(0,\hat{r}_{B}) > S_{A}(0,\hat{r}_{A})$.
% Let denote $A_{1}$ as a state of a system $A$. 
We are considering an adiabatic transition $A \rightarrow B$ starting from a state $A_{1}$ of system $A$ at a nonzero $T_{1}$, whose entropy $S_{A}(T_{1},\hat{r}_{A})$ is less than $S_{B}(0,\hat{r}_{B})$, and asking the possibility of reaching $T=0$ in system $B$.
% , whose entropy is $S_{B}(0,\hat{r}_{B})$. 

A transition from $S_{A}(T_{1},\hat{r}_{A})$ to $S_{B}(0,\hat{r}_{B})$ is compatible with the second law, because of increase in $S$. However, the internal constraint strictly restricts the accessible regions $(T,S)$. The region allowed for $A$ is disconnected from the region allowed for $B$, as indicated by the hatched area in Fig.~\ref{fig:TS-diagram}(b).
In terms of the frozen coordinate $r$, the shaded area corresponds to a range of $r$ from $\hat{r}_{A}$ to $\hat{r}_{B}$. 
% When the transition crosses the hatched region, $\hat{r}$ jumps from $\hat{r}_{A}$ to $\hat{r}_{B}$. 
%%%% 29 March 2018
The adiabatic process starting from $A_{1}$ proceeds until reaching a state $A_{2}$ at the boundary of $A$. 
At this point, the internal constraint is removed. Then, the frozen coordinate $\hat{r}$ becomes a real variable $r$. The system does change from $A$ to $B$. Since, during changing of $r$ from $\hat{r}_{A}$ to $\hat{r}_{B}$, the system is in nonequilibrium, the entropy jumps by $\Delta S^{AB}(T_{2}) = S_{B}(T_{2}, \hat{r}_{B} ) -S_{A}(T_{2}, \hat{r}_{A})$. 
Here, $T$ is assumed to be constant. Usually in solids, the temperature increases, $T_{2'} > T_{2}$, except a negligible decrease due to the volume expansion if presents (see Appendix). The constant $T$ is, therefore, the best case.
% There are cases of decreasing $T$ for such gases having negative Joule-Thomson coefficient. But, this is only a second-order effect, the decrease $\Delta T$ is smaller than $T_{2}$. The following argument is valid as long as $T_{2'} > 0$. 
At the end point $B_{2'}$, system $B$ attains an entropy value $S_{B}(2')$, which is larger than $S_{B}(0)$. From $B_{2'}$, there is no adiabatic path to the lowest-entropy state $B_{0}$, as is evident from Fig~\ref{fig:TS-diagram}(b). Therefore, we cannot reach $B_{0}$ from $A_{1}$ all the more.
The residual entropy $S_{0}^{AB}$ is the limiting value of the discontinuity $\Delta S^{AB}(T_{2})$ as $T_{2} \rightarrow 0$. 

%
% A lessen of the above argument is important for deep understanding of entropy.
The above argument of proof is instructive for understanding of the accessibility of thermodynamic states.
Reversible processes are certainly the processes retaining a constant entropy, $\Delta S_{12}=0$. However, the converse is not always true. When the end states 1 and 2 are separated by an internal constraint, we cannot find adiabatic paths with a constant $S$ from 1 to 2. On the way from 1 to 2, an entropy jump occurs. The origin of entropy has been changed. In this way, judging the accessibility by comparing only the end states is risky.
% We need to track the change of constraints during the process. This is why we have employed a roundabout way of the proof above.
%

% At last, a few comments are worthwhile to be mentioned. The argument of adiabatic accessibility, though elegant, there is a pitfall behind the brief expression. The adiabatic accessibility is expressed as, $A(\alpha,T_{1}) \prec B(\beta, 0)$ if $S_{A}(\alpha,T_{1}) \leq S_{B}(\beta, 0)$ \cite{Lieb99}. In this expression, we compare only the terminal points, $A(\alpha,T_{1})$ and $B(\beta, 0)$ \cite{Wreszinski09}. When there is a residual entropy $S_{0}^{AB}$ between $A$ and $B$, even if $S_{A}(\alpha,T_{1}) = S_{B}(\beta, 0)$, two states are not necessarily connected by a reversible process. 
% $A(\alpha,T_{1}) \not\prec B(\beta, 0)$

% It is intuitively evident the residual entropy cannot be used for attaining zero temperature.
% misleading arguments for residual entropy exist.
There are confusing descriptions in the literature.
Wilks ascribed the impossibility of reaching zero temperature when residual entropy exists to the impossibility of changing frozen-in states by external parameter (\cite{Wilks}, p.~115). 
This is not true. As described in Paper (I), there must be at least one reversible path for any transition $A \rightarrow B$, and this is so even when the transition is caused by removing the internal constraint.
% A residual entropy $S_{0}^{AB}$ stems from a transition $A \rightarrow B$, where two systems are separated by internal constraints. 
Removing the internal constraint alone is irreversible. However, if we bring another system $C$, and make a thermal contact to the combined system $A+B$, we will find a reversible path. In fact, the measurement of residual entropy $S_{0}^{AB}$ is carried out in this way. However, the fact of the increase in entropy from $S_{A}(2)$ to $S_{B}(2')$ does no change, because entropy is a state variable. 
By replacing the irreversible path $2 \rightarrow 2'$ with a reversible one, the increase $\Delta S^{AB}(T_{2})$ must be compensated with a decrease in entropy of $C$. This means that heat is injected to $A$, and therefore it is not usable for cooling. 

% A random alloy A$_{1-x}$B$_{x}$ has a mixing entropy, which remains at $T=0$ as the residual entropy $S_{0}^{AB}$. Starting from a separated metals, A and B, at $T=0$, we can obtain the state of alloy at $T=0$ by reversible process. First, heating A and B quasistatically, and melt them separately. In the liquid phase, we can mix two liquid metals by using a semipermeable membrane, and finally bring the obtained alloy back to the state of $T=0$.

% We can perform all the process adiabatically, {\it e.g.}, by expansion and compression. We cannot achieve the zero temperature by this process. Although applying these processes to solids is very inefficient, it is enough to show the existence in principle. 
% From the crystalline phase of a glass, we can obtain the glass state, by first thawing the crystal and then cooling the melt slowly. Such a transition may be regarded as a reversible process. The specific heat of the glass is a well-defined quantity, and the value of entropy at $T=0$ is independent of the cooling rate.
% It is impossible to answer all these questions in one paper. If these questions are not covered by the arguments given in this paper, it should be left in future study.

When looking at specific problems of condensed matters, we find many unsolved or tricky arguments on the third law, for which deep knowledge of individual materials is required. These topics spread from ideal gases \cite{Boyer70}, Bose-Einstein gases \cite{Wheeler91,Wheeler92,Lingua18}, quantum cooling, \cite{Cleuren12, Levy12,Sordal16,Kirkpatrick15}, and even to black holes \cite{Wald97}.
Before going to individual problems, we have to know exactly what is the third law. This paper gives general feature of the third law from the macroscopic viewpoint. 
% These topics are beyond the scope of this paper. 

%%%%%%%%%%%%%%%%%%%%%%%%%%%%%%%%%%%%%%%%%%
% \section{Conclusions}
%%%%%%%%%%%%%%%%%%%%%%%%%%%%%%%%%%%%%%%%%%%%%%%%%%

\begin{acknowledgments}
The author thanks J.~C.~Wheeler and S.~Fujita for useful discussions.
\end{acknowledgments}

\appendix
\section{Appendix}
In this appendex, we show that, when the internal constraint is adiabatically removed from system $A$ (the process $2 \rightarrow 2'$ in Fig.~\ref{fig:TS-diagram}(b)), the temperature of $A$ is increased or at best constant. %, except minor secondary effects.
% The proof is not in a completely general manner but rather in an enumerating manner. However, this practically exhausts all the conceivable cases. 
In the following, we show this by separately treating the cases in the relationships between the internal energies $U_{A}(2)$ and $U_{B}(2')$.
All the reference numbers for figures and Examples are those of Paper I.

\noindent
{\bf Case 1}, $U_{B}(2') > U_{A}(2)$

This is the case of Example 6 of creating defects in silicon by an electron irradiation.
Interstitial atoms are in excited states with the total energy increase by $\Delta U_{I}$. 
Introducing $n_{I}$ interstitial atoms among $N_{I}$ available interstitial sites yields an entropy increase $\Delta S_{\rm dis} = \ln (N_{I}/n_{I})$. Although system $A$ adiabatically increases the entropy $S_{A}$ by $\Delta S_{\rm dis}$, the process does not occur until mechanical work $W$ is supplied in order to increase the total energy by $\Delta U_{I}$. This $W$ must come from the energy transfer from the incident electron beam. The best case is the perfect energy transfer $W = \Delta U_{I}$. % This does not conflict with the second law, despite that the increase $\Delta S_{\rm dis}$ seems to inhibit $W = \Delta U_{I}$. 
% Although the energy required to displace atoms is supplied by the mechanical work of electrons, the efficiency of displacement is very small. 
As in Example 2, if we graspe perfectly atom positions, and if we can move each atom to the intended position exactly, the entropy change $\Delta S$ must be 0 but not $\Delta S_{\rm dis}$. However, this occurs only for the case of head-on collision. Only a tiny incline of the incident beam causes a chain of uncontrollable collisions. Most part of the supplied mechanical work is dissipated into a heat generation $Q_{\rm gen}$. Accordingly, $\Delta U_{I} = W - Q_{\rm gen}$, and the total entropy is further increased by $Q_{\rm gen}/T_{2}$. This results in an increase in $T_{2'}$.

\noindent
{\bf Case 2}, $U_{B}(2') = U_{A}(2)$

This is the case of Example 5 of formation a random alloy $XY$. Here, $A$ is the separated system $(X|Y)$ and $B$ is the mixed system $(XY)$.
In this case, no external work is necessary. The natural process occurring at a finite temperature $T_{2}$ is intermixing of $X$ and $Y$. There is no reason for decreasing in $T$.

\noindent
{\bf Case 3}, $U_{B}(2') < U_{A}(2)$

A rare case is that a disordered system $B$ has a lower energy than the ordered system $A$. The present theory does treat even this case (see Sec.~III B of Paper I), although traditional theories exclude it.
Because the process is adiabatic, the decrease in $U$ turns to an internal generation of heat $Q$. The decrease in $T$ never takes place.

%%%%%%%%%%%%%%%%%%%%%%%%%%%%%%%%%%%%%%%%%%

%=====================================
% References, variant A: internal bibliography
%=====================================
% \reftitle{References}
%\begin{thebibliography}{999}
% Reference 1
%Author1, T. The title of the cited article. {\em Journal Abbreviation} {\bf 2008}, {\em 10}, 142-149, DOI.

% \bibliography{thermo}

\begin{thebibliography}{34}
\expandafter\ifx\csname natexlab\endcsname\relax\def\natexlab#1{#1}\fi
\expandafter\ifx\csname bibnamefont\endcsname\relax
  \def\bibnamefont#1{#1}\fi
\expandafter\ifx\csname bibfnamefont\endcsname\relax
  \def\bibfnamefont#1{#1}\fi
\expandafter\ifx\csname citenamefont\endcsname\relax
  \def\citenamefont#1{#1}\fi
\expandafter\ifx\csname url\endcsname\relax
  \def\url#1{\texttt{#1}}\fi
\expandafter\ifx\csname urlprefix\endcsname\relax\def\urlprefix{URL }\fi
\providecommand{\bibinfo}[2]{#2}
\providecommand{\eprint}[2][]{\url{#2}}

\bibitem[{\citenamefont{Beattie and Oppenheim}(1979)}]{Beattie}
\bibinfo{author}{\bibfnamefont{J.~A.} \bibnamefont{Beattie}} \bibnamefont{and}
  \bibinfo{author}{\bibfnamefont{I.}~\bibnamefont{Oppenheim}},
  \emph{\bibinfo{title}{Principles of Thermodynamics}}
  (\bibinfo{publisher}{Elsevier}, \bibinfo{address}{Amsterdam},
  \bibinfo{year}{1979}).

\bibitem[{\citenamefont{Callen}(1985)}]{Callen}
\bibinfo{author}{\bibfnamefont{H.}~\bibnamefont{Callen}},
  \emph{\bibinfo{title}{Thermodynamics and an Introduction to Thermostatistics,
  2nd ed.}} (\bibinfo{publisher}{Wiley}, \bibinfo{address}{New York},
  \bibinfo{year}{1985}).

\bibitem[{com()}]{comment3}
\bibinfo{note}{In the literature, it is often seen that an additional
  restriction ``in a finite number of operations" is imposed to Expression
  (II). This restriction is unnecessary \cite{Buchdahl}. Single operation is
  not well-defined. An infinite number of operations may be operationally
  attainable if each operation is infinitesimal. (\cite{Hatsopoulos}, p.~$30$
  in Forward).}

\bibitem[{\citenamefont{Hatsopoulos and Keenan}(1965)}]{Hatsopoulos}
\bibinfo{author}{\bibfnamefont{G.~N.} \bibnamefont{Hatsopoulos}}
  \bibnamefont{and} \bibinfo{author}{\bibfnamefont{J.~H.}
  \bibnamefont{Keenan}}, \emph{\bibinfo{title}{Principles of General
  Thermodynamics}} (\bibinfo{publisher}{John Wiley \& Sons, Inc.},
  \bibinfo{address}{New York}, \bibinfo{year}{1965}).

\bibitem[{\citenamefont{Hasse}(1971)}]{Hasse}
\bibinfo{author}{\bibfnamefont{R.}~\bibnamefont{Hasse}},
  \emph{\bibinfo{title}{Physical Chemistry: An Advanced Treatise, Vol. 1/ Thermodynamics}},
  \bibinfo{title}{ed. W. Jost, Chap.~1}, 
   (\bibinfo{publisher}{Academic},
  \bibinfo{address}{New York}, \bibinfo{year}{1971}).

\bibitem[{\citenamefont{Levine}(1983)}]{Levine}
\bibinfo{author}{\bibfnamefont{I.~N.} \bibnamefont{Levine}},
  \emph{\bibinfo{title}{Physical Chemistry, 2nd ed.}}
  (\bibinfo{publisher}{McGraw-Hill}, \bibinfo{address}{New York},
  \bibinfo{year}{1983}).

\bibitem[{\citenamefont{Landsberg}(1957)}]{Landsberg57}
\bibinfo{author}{\bibfnamefont{P.~T.} \bibnamefont{Landsberg}},
  \bibinfo{journal}{Rev. Mod. Phys.} \textbf{\bibinfo{volume}{28}},
  \bibinfo{pages}{363} (\bibinfo{year}{1957}).

\bibitem[{\citenamefont{Landsberg}(1978)}]{Landsberg78}
\bibinfo{author}{\bibfnamefont{P.~T.} \bibnamefont{Landsberg}},
  \emph{\bibinfo{title}{Thermodynamics and Statistical Mechanics}}
  (\bibinfo{publisher}{Oxford}, \bibinfo{address}{Oxford},
  \bibinfo{year}{1978}).

\bibitem[{\citenamefont{Landsberg}(1997)}]{Landsberg97}
\bibinfo{author}{\bibfnamefont{P.~T.} \bibnamefont{Landsberg}},
  \bibinfo{journal}{Am. J. Phys.} \textbf{\bibinfo{volume}{65}},
  \bibinfo{pages}{269} (\bibinfo{year}{1997}).

\bibitem[{\citenamefont{III et~al.}(2017)\citenamefont{III, Fujita, and
  Suzuki}}]{McNabb17}
\bibinfo{author}{\bibfnamefont{J.~R.~McNabb~} \bibnamefont{III}},
  \bibinfo{author}{\bibfnamefont{S.}~\bibnamefont{Fujita}}, \bibnamefont{and}
  \bibinfo{author}{\bibfnamefont{A.}~\bibnamefont{Suzuki}},
  \bibinfo{journal}{J. Mod. Phys.} \textbf{\bibinfo{volume}{8}},
  \bibinfo{pages}{839} (\bibinfo{year}{2017}).

\bibitem[{\citenamefont{Masanes and Oppenheim}(2017)}]{Masanes17}
\bibinfo{author}{\bibfnamefont{L.}~\bibnamefont{Masanes}} \bibnamefont{and}
  \bibinfo{author}{\bibfnamefont{J.}~\bibnamefont{Oppenheim}},
  \bibinfo{journal}{Nature Commun.} \textbf{\bibinfo{volume}{8}},
  \bibinfo{pages}{14538} (\bibinfo{year}{2017}).

\bibitem[{\citenamefont{Simon}(1951)}]{Simon51}
\bibinfo{author}{\bibfnamefont{F.~E.} \bibnamefont{Simon}},
  \bibinfo{journal}{Zeit. f. Naturforsh.} \textbf{\bibinfo{volume}{6a}},
  \bibinfo{pages}{397} (\bibinfo{year}{1951}).

\bibitem[{\citenamefont{Simon}(1927)}]{Simon27}
\bibinfo{author}{\bibfnamefont{F.}~\bibnamefont{Simon}},
  \bibinfo{journal}{Zeit. f. Phys.} \textbf{\bibinfo{volume}{41}},
  \bibinfo{pages}{806} (\bibinfo{year}{1927}).

\bibitem[{\citenamefont{Fowler and Guggenheim}(1952)}]{Fowler-Guggenheim}
\bibinfo{author}{\bibfnamefont{R.}~\bibnamefont{Fowler}} \bibnamefont{and}
  \bibinfo{author}{\bibfnamefont{E.~A.} \bibnamefont{Guggenheim}},
  \emph{\bibinfo{title}{Statistical Thermodynamics, 3rd ed.}}
  (\bibinfo{publisher}{Cambridge}, \bibinfo{address}{London},
  \bibinfo{year}{1952}).

\bibitem[{\citenamefont{Epstein}(1961)}]{Epstein}
\bibinfo{author}{\bibfnamefont{P.~S.} \bibnamefont{Epstein}},
  \emph{\bibinfo{title}{Textbook of Thermodynamics, 8th ed.}}
  (\bibinfo{publisher}{Wiley}, \bibinfo{address}{New York},
  \bibinfo{year}{1961}).

\bibitem[{\citenamefont{Pippard}(1957)}]{Pippard}
\bibinfo{author}{\bibfnamefont{A.~B.} \bibnamefont{Pippard}},
  \emph{\bibinfo{title}{Elements of Classical Thermodynamics}}
  (\bibinfo{publisher}{Cambridge}, \bibinfo{address}{Cambridge},
  \bibinfo{year}{1957}).

\bibitem[{adi()}]{adiabatic-change}
\bibinfo{note}{By the word adiabatic change of a system $A$, it is meant that
  there is no heat exchange between $A$ and its surroundings but work
  interactions between them are allowed \cite{Tasaki}.}

\bibitem[{\citenamefont{Lieb and Yngvason}(1999)}]{Lieb99}
\bibinfo{author}{\bibfnamefont{E.~H.} \bibnamefont{Lieb}} \bibnamefont{and}
  \bibinfo{author}{\bibfnamefont{J.}~\bibnamefont{Yngvason}},
  \bibinfo{journal}{Phys. Rep.} \textbf{\bibinfo{volume}{310}},
  \bibinfo{pages}{1} (\bibinfo{year}{1999}).

\bibitem[{\citenamefont{Wreszinski and Abdalla}(2009)}]{Wreszinski09}
\bibinfo{author}{\bibfnamefont{W.~F.} \bibnamefont{Wreszinski}}
  \bibnamefont{and} \bibinfo{author}{\bibfnamefont{E.}~\bibnamefont{Abdalla}},
  \bibinfo{journal}{J. Stat. Phys.} \textbf{\bibinfo{volume}{134}},
  \bibinfo{pages}{781} (\bibinfo{year}{2009}).

\bibitem[{\citenamefont{Guggenheim}(1967)}]{Guggenheim}
\bibinfo{author}{\bibfnamefont{E.~A.} \bibnamefont{Guggenheim}},
  \emph{\bibinfo{title}{Thermodynamics and Advanced Treatment for Chemists and
  Physicists, 5th ed.}} (\bibinfo{publisher}{North-Holland},
  \bibinfo{address}{Amsterdam}, \bibinfo{year}{1967}).

\bibitem[{\citenamefont{Wilson}(1957)}]{Wilson}
\bibinfo{author}{\bibfnamefont{A.~H.} \bibnamefont{Wilson}},
  \emph{\bibinfo{title}{Thermodynamics and Statistical Thermodynamics}}
  (\bibinfo{publisher}{Cambridge}, \bibinfo{address}{Cambridge},
  \bibinfo{year}{1957}).

\bibitem[{\citenamefont{Wilks}(1961)}]{Wilks}
\bibinfo{author}{\bibfnamefont{J.}~\bibnamefont{Wilks}},
  \emph{\bibinfo{title}{The Third Law of Thermodynamics}}
  (\bibinfo{publisher}{Oxford}, \bibinfo{address}{London},
  \bibinfo{year}{1961}).

\bibitem[{\citenamefont{Gyftopoulos and Beretta}(2005)}]{Gyftopoulos}
\bibinfo{author}{\bibfnamefont{E.~P.} \bibnamefont{Gyftopoulos}}
  \bibnamefont{and} \bibinfo{author}{\bibfnamefont{G.~P.}
  \bibnamefont{Beretta}}, \emph{\bibinfo{title}{Thermodynamics - Foundations
  and Applications}} (\bibinfo{publisher}{Dover Pub.}, \bibinfo{address}{New
  York}, \bibinfo{year}{2005}).

\bibitem[{\citenamefont{Boyer}(1970)}]{Boyer70}
\bibinfo{author}{\bibfnamefont{T.~H.} \bibnamefont{Boyer}},
  \bibinfo{journal}{Phys. Rev. D} \textbf{\bibinfo{volume}{1}},
  \bibinfo{pages}{1526} (\bibinfo{year}{1970}).

\bibitem[{\citenamefont{Wheeler}(1991)}]{Wheeler91}
\bibinfo{author}{\bibfnamefont{J.~C.} \bibnamefont{Wheeler}},
  \bibinfo{journal}{Phys. Rev. A} \textbf{\bibinfo{volume}{43}},
  \bibinfo{pages}{5289} (\bibinfo{year}{1991}).

\bibitem[{\citenamefont{Wheeler}(1992)}]{Wheeler92}
\bibinfo{author}{\bibfnamefont{J.~C.} \bibnamefont{Wheeler}},
  \bibinfo{journal}{Phys. Rev. A} \textbf{\bibinfo{volume}{45}},
  \bibinfo{pages}{2637} (\bibinfo{year}{1992}).

\bibitem[{\citenamefont{Lingua et~al.}(2018)\citenamefont{Lingua, Richaud, and
  Penna}}]{Lingua18}
\bibinfo{author}{\bibfnamefont{F.}~\bibnamefont{Lingua}},
  \bibinfo{author}{\bibfnamefont{A.}~\bibnamefont{Richaud}}, \bibnamefont{and}
  \bibinfo{author}{\bibfnamefont{V.}~\bibnamefont{Penna}},
  \bibinfo{journal}{Entropy} \textbf{\bibinfo{volume}{20}}, \bibinfo{pages}{84}
  (\bibinfo{year}{2018}).

\bibitem[{\citenamefont{Cleuren et~al.}(2012)\citenamefont{Cleuren, Rutten, and
  den Broeck}}]{Cleuren12}
\bibinfo{author}{\bibfnamefont{B.}~\bibnamefont{Cleuren}},
  \bibinfo{author}{\bibfnamefont{B.}~\bibnamefont{Rutten}}, \bibnamefont{and}
  \bibinfo{author}{\bibfnamefont{C.~V.} \bibnamefont{den Broeck}},
  \bibinfo{journal}{Phys. Rev. Lett.} \textbf{\bibinfo{volume}{108}},
  \bibinfo{pages}{120603} (\bibinfo{year}{2012}).

\bibitem[{\citenamefont{Levy et~al.}(2012)\citenamefont{Levy, Alicki, and
  Kosloff}}]{Levy12}
\bibinfo{author}{\bibfnamefont{A.}~\bibnamefont{Levy}},
  \bibinfo{author}{\bibfnamefont{R.}~\bibnamefont{Alicki}}, \bibnamefont{and}
  \bibinfo{author}{\bibfnamefont{R.}~\bibnamefont{Kosloff}},
  \bibinfo{journal}{Phys. Rev. E} \textbf{\bibinfo{volume}{85}},
  \bibinfo{pages}{061126} (\bibinfo{year}{2012}).

\bibitem[{\citenamefont{Sordal et~al.}(2016)\citenamefont{Sordal, Berglie, and
  Galperin}}]{Sordal16}
\bibinfo{author}{\bibfnamefont{V.~B.} \bibnamefont{Sordal}},
  \bibinfo{author}{\bibfnamefont{J.}~\bibnamefont{Berglie}}, \bibnamefont{and}
  \bibinfo{author}{\bibfnamefont{Y.~M.} \bibnamefont{Galperin}},
  \bibinfo{journal}{Phys. Rev. E} \textbf{\bibinfo{volume}{93}},
  \bibinfo{pages}{032102} (\bibinfo{year}{2016}).

\bibitem[{\citenamefont{Kirkpatrick and Belitz}(2015)}]{Kirkpatrick15}
\bibinfo{author}{\bibfnamefont{T.~R.} \bibnamefont{Kirkpatrick}}
  \bibnamefont{and} \bibinfo{author}{\bibfnamefont{D.}~\bibnamefont{Belitz}},
  \bibinfo{journal}{Phys. Rev. Lett.} \textbf{\bibinfo{volume}{115}},
  \bibinfo{pages}{020402} (\bibinfo{year}{2015}).

\bibitem[{\citenamefont{Wald}(1997)}]{Wald97}
\bibinfo{author}{\bibfnamefont{R.~M.} \bibnamefont{Wald}},
  \bibinfo{journal}{Phys. Rev. D} \textbf{\bibinfo{volume}{56}},
  \bibinfo{pages}{6467} (\bibinfo{year}{1997}).

\bibitem[{\citenamefont{Buchdahl}(1966)}]{Buchdahl}
\bibinfo{author}{\bibfnamefont{H.~A.} \bibnamefont{Buchdahl}},
  \emph{\bibinfo{title}{The Concepts of Classical Thermodynamics}}
  (\bibinfo{publisher}{Cambridge}, \bibinfo{address}{Cambridge},
  \bibinfo{year}{1966}).

\bibitem[{\citenamefont{Tasaki}(2000)}]{Tasaki}
\bibinfo{author}{\bibfnamefont{H.}~\bibnamefont{Tasaki}},
  \emph{\bibinfo{title}{Thermodynamics: From the Contemporary Viewpoint}}
  (\bibinfo{publisher}{Baifukan}, \bibinfo{address}{Tokyo},
  \bibinfo{year}{2000}), \bibinfo{note}{in Japanese}.

\end{thebibliography}

%%%%%%%%%%%%%%%%%%%%%%%%%%%%%%%%%%%%%%%%%%

%%%%%%%%%%%%%%%%%%%%%%%%%%%%%%%%%%%%%%%%%%
\end{document}